# Induction-Detection Electron Spin Resonance with Sensitivity of 1000 Spins:

# En Route to Scalable Quantum Computations

Aharon Blank,[1] Ekaterina Dikarov, Roman Shklyar, and Ygal Twig

Schulich Faculty of Chemistry
Technion – Israel Institute of Technology
Haifa, 32000
Israel

[1] Corresponding author contact details: Aharon Blank, Schulich Faculty of Chemistry, Technion – Israel Institute of Technology, Haifa, 32000, Israel, phone: +972-4-829-3679, fax: +972-4-829-5948, e-mail: ab359@tx.technion.ac.il.

**Spin-based quantum computation (QC) in the solid state is considered to be one of the most promising approaches to scalable quantum computers[1-4]. However, it faces problems such as initializing the spins, selectively addressing and manipulating single spins, and reading out the state of the individual spins. We have recently sketched a scheme that potentially solves all of these problems[5]. This is achieved by making use of a unique phosphorus-doped $^{28}$Si sample ($^{28}$Si:P), and applying powerful new electron spin resonance (ESR) techniques for parallel excitation, detection, and imaging in order to implement QCs and efficiently obtain their results. The beauty of our proposed scheme is that, contrary to other approaches, single-spin detection sensitivity is not required and a capability to measure signals of ~100-1000 spins is sufficient to implement it. Here we take the first experimental step towards the actual implementation of such scheme. We show that, by making use of the smallest ESR resonator constructed to date (~5 μm), together with a unique cryogenic amplification scheme and sub-micron imaging capabilities, a sensitivity of less than 1000 electron spin is obtained with spatial resolution of ~500 nm. This is the most sensitive induction-detection experiment carried out to date, and such capabilities put this approach on the fast track to the demonstration of a scalable QC capability.**

In order to realize a scalable quantum computer it is necessary to comply with the so-called "DiVincenzo criteria"[6], which include: (i) having a scalable physical system with well-characterized qubits; (ii) the ability to initialize the state of the qubits to a simple fiducial state such as |000 . . .>; (iii) long coherence times, much longer than the gate operation time; (iv) a ''universal'' set of quantum gates; and (v) a qubit-specific measurement capability. A scheme for addressing all these issues is briefly presented in Fig. 1[5]. It is based on the use of a uniquely fabricated $^{28}$Si:P array, coupled with high static magnetic field gradients and fast switchable high polarizing fields, along with ultra-sensitive induction-detection ESR. The use of the electron and nuclear spins in $^{28}$Si:P as qubits is known to be very promising in the context of QCs[7]. The long coherence time of the electrons and nuclei in this system (in the range of seconds[8]) compared to the short interaction and manipulation times of the electron spins (in the range of 10-1000 ns), puts it on a par with the most advanced ideas for QCs. In addition, the electrons' spin-lattice relaxation time of $^{28}$Si:P can be effectively controlled by means of light, making it possible to greatly increase spin polarization using an appropriate short pulse of large static field[5]. The fabrication of such an array is somewhat beyond the current capabilities of nanotechnology, but not very far[9]. Apart from the unique sample itself, the major missing component that is needed for the realization of the proposed scheme is having an induction-detection (sometimes called Faraday detection) capability to detect the signal from only ~100-1000 electron spins (in a reasonable averaging time of ~1 hour). Thus, while single-spin sensitivity is not required to operate such QCs, these values are still very far from the capabilities of the best commercial ESR systems whose maximum levels of sensitivity are only ~$10^8$-$10^9$ spins[10], and thus seem to make the proposed scheme unattainable.

The absolute spin sensitivity in induction detection ESR is proportional to $1/\sqrt{V_c}$, where $V_c$ is the effective volume of the resonator employed[11]. Commercial systems commonly employ relatively large resonators which, at a typical frequency of ~10 GHz, have an effective volume of a few milliliters down to a few microliters at most. Recently, extensive work has been carried out with the aim of designing and constructing resonators with much smaller effective volumes while maintaining reasonably high quality ($Q$) factors[11-13]. The latest of these efforts is our continued work on a set of so-called surface loop-gap microresonators that have a very small internal diameter, reaching just 5 μm ($V_c \approx 0.1$ nl – see Fig. 1) in our most recent designs for operation at the Ku microwave band (~15-17 GHz)[14]. This resonator exhibited a measured spin sensitivity of ~$3\times10^7$ spins/√Hz (or ~$5\times10^5$ spins for 1 hour of averaging) at 15.76 GHz for a sample of γ-irradiated $SiO_2$, measured at room temperature[14]. In parallel, we have also recently measured a sample of $^{28}$Si:P at a temperature of 10 K using a slightly larger resonator featuring a 20-μm internal diameter, which provided spin sensitivity of ~4000 spins (with signal-to-noise-ratio (SNR) of 1) for 2 hours of acquisition time [15].

In the present work we undertook a significant experimental step towards achieving the sensitivity threshold that would allow the implementation of our proposed QC scheme by means of induction detection. The recent measurements were also carried out with a $^{28}$Si:P sample at a temperature of 9.5 K with the 20-μm and the 5-μm resonators, which are the smallest of their kind. The use of such small resonators at cryogenic temperatures enabled us to improve on the sensitivity obtained from our previous measurements. Furthermore, additional significant improvements in spin sensitivity for both resonators were obtained through the use of a newly fabricated cryogenic probe (Fig. S1) which incorporates a cryogenic low noise amplifier (model LNF-LNC6_20A from Low Noise Factory AB, Sweden), and a cryogenic magnetically-shielded circulator (model PTG1218KCSZ from QuinStar Technology Inc., USA). In the new microwave configuration, the pulsed microwave excitation signal goes first through the circulator, then reaches the resonator and returns to the circulator and to the cryogenic amplifier. Since both the circulator and the amplifier are cooled to ~10 K, the noise in the detection system is decreased by a factor of ~5 compared to the use of an external circulator and amplifier. It should be noted that such cryogenic low noise amplifiers are very sensitive to the applied microwave power and a level of more than ~1 mW would damage them. Nevertheless, we can manage with this limitation without the use of a protection switch before the cryogenic amplifier (which would greatly deteriorate its noise performance) because we make use of a surface resonator for which a power level of ~0.5 mW is more than enough to efficiently excite the spins in the sample in pulsed ESR[14]. The probe itself has also microimaging capabilities and is a greatly improved version of the one described in ref [15].

The results of our experiments with a 10-μm-thick $^{28}$Si:P sample containing $10^{16}$ P atoms in 1 cm$^3$ (described in ref [16]) are shown in Figs. 2-3. The sample is placed face down on the resonator (Fig. S2). Our home-made pulsed ESR imaging system is described in ref[11]. For signal acquisition, we employed a Carr-Purcell-Meiboom-Gill (CPMG) pulse sequence with a repetition rate of 1000 Hz, π/2-π pulse separation, $\tau$ = 1.2 μs, and a data acquisition window of 1 μs. Figure 2 shows the acquired echo signal compared to the noise level for an averaging time of 1 sec, i.e., for 1,000 CPMG trains (data was also averaged along each 160-π-pulse CPMG train). The measured SNR was ~1486 (~581) for the 20-μm (5-μm) resonators. The number of spins in the effective volume of these two resonators can be estimated to be not more than ~4.8×10$^7$ (~2.25×10$^7$), based on the calculated volume of the resonator from which most of the signal is acquired ~60×20×4 (5×150×3) [μm$^3$][14]. This provides an initial estimate of spin sensitivity (for SNR=1) of ~3.2×10$^4$ (~3.87×10$^4$) spins/√Hz for the 20-μm (5-μm) resonators employed here.

At first glance it seems that the use of the smaller resonator did not increase, and even reduced, the spin sensitivity, while in theory it should have improved the latter by a factor of ~1.62 compared to the larger resonator[14]. However, our calculations above, based on the echo signal, assumed constant ESR sensitivity throughout the resonator's effective volume, while in practice some parts are more sensitive than others (and these are the places where it would be preferable to place small samples)[14]. Thus, in order to provide an answer to this issue and, mainly, to offer more exact measured values for the spin sensitivity of our resonators *at their most sensitive spot*, we acquired two dimensional high resolution images of the sample in the resonators, as shown in Fig. 3. These imaging results also demonstrate our ability to obtain a very high spatial resolution with this type of sample – which is of relevance for our proposed QC scheme. Although the sample is positioned so that it covers the resonator's entire central area (Fig. S2), the signal originates only from areas where the resonator has a strong microwave magnetic field component ($B_1$) and the image should correspond to the calculated spatial distribution of $B_1^2$ [13]. For the 20-μm (5-μm) resonator the size of each voxel in this image is 0.5×0.75 (1×1.2) μm, and although the sample's thickness is 10 μm most of the signal originates only at the first 5 (3) μm above the resonator's surface , due to the fast decay of $B_1$ when going out of plane[14,15]. Thus, the imaging experiment, combined with the out-of-plane field's calculated data, provides us with the voxel volume of 1.87 (3.6) μm$^3$, which contain 1.87×10$^4$ (3.6×10$^4$) spins. The maximum signal-to-noise ratio (measured at the image's peripheral parts) is found to be 78 (97), giving a spin sensitivity (with SNR=1) of 240 (370) spins for the total measurement time employed for image acquisition (10 hours for both images). In other words, based on the ESR imaging results, spin sensitivity is found to be 240×√(3600×10)≈4.5×10$^4$ (370×√(3600×10)≈7.0×10$^4$) spins/√Hz, for the 20-μm (5-μm) resonators employed here, which is somewhat worse than our above estimate formulated on the bases of the 1-sec-echo data acquisition. It should be noted, however, that the limited stability of our system may explain the reduction in performance during the prolonged imaging data acquisition period (used for averaging).

The sensitivity obtained in the present measurements corresponds well to the theoretical predictions for spin sensitivity of ~$8.6\times10^4$ (~$5.3\times10^4$) spins/√Hz, for the 20-μm (5-μm) resonators[14]. Although the 5-μm resonator gave lower experimental sensitivity than the 20-μm and is further apart from the theoretical prediction, we believe that this is due to our limited sample-resonator attachment capability. This means that something, probably some duct, prevented us from placing the sample right on top of the resonator, and this has a greater effect on the signal at the 5-μm resonator than at the larger one. Thus, it is highly plausible that the theoretical values do represent that which can be achieved with more direct sample/resonator coupling, where the 5 μm resonator would be the most sensitive one. The experimental results for spin sensitivity demonstrated here are by far the best obtained to date with induction-detection ESR and can support many important future experiments with spin-limited paramagnetic materials (e.g. defects and impurities in semiconductors and spin-labeled biological molecules), including, as noted above, the demonstration of a unique scalable QC scheme. It should be emphasized that while alternative detection methods, such as magnetic resonance force microscopy[17] and indirect detection via Nitrogen Vacancy centers in diamonds,[18] exhibited in the past much better sensitivity and spatial resolution, they are highly limited and as such are not useful yet to the above and similar applications. Finally, spin sensitivity can be further improved by a factor of up to ~25 by going to higher static fields (3.4 T) and smaller resonators (down to ~1 μm), as we have recently outlined in details[14].

## Acknowledgments

This work was partially supported by grant # 213/09 from the ISF, grant #201665 from the ERC, grant # G-1032-18.14/2009 from the GIF, and by RBNI and MNFU at the Technion.

## Contributions

A.B. initiated and supervised the research, and was involved in the resonator and probe design; E.D. prepared the microresonators and performed the experiments; R.S. performed the data analysis for the images, and Y.T designed and tested the resonators and MW system.

## References

1   Kane, B. E. A silicon-based nuclear spin quantum computer. *Nature* **393**, 133-137 (1998).
2   Harneit, W., Meyer, C., Weidinger, A., Suter, D. & Twamley, J. Architectures for a spin quantum computer based on endohedral fullerenes. *Physica Status Solidi B-Basic Research* **233**, 453-461 (2002).
3   Cerletti, V., Coish, W. A., Gywat, O. & Loss, D. Recipes for spin-based quantum computing. *Nanotechnology* **16**, R27-R49, doi:Doi 10.1088/0957-4484/16/4/R01 (2005).
4   Ju, C. Y., Suter, D. & Du, J. F. An endohedral fullerene-based nuclear spin quantum computer. *Phys. Lett. A* **375**, 1441-1444, doi:DOI 10.1016/j.physleta.2011.02.031 (2011).
5   Blank, A. Scheme for a spin-based quantum computer employing induction detection and imaging. *arXiv:1302.1653 [quant-ph]*.
6   DiVincenzo, D. P. The physical implementation of quantum computation. *Fortschr Phys* **48**, 771-783 (2000).
7   Morton, J. J. L., McCamey, D. R., Eriksson, M. A. & Lyon, S. A. Embracing the quantum limit in silicon computing. *Nature* **479**, 345-353, doi:Doi 10.1038/Nature10681 (2011).
8   Tyryshkin, A. M. *et al.* Electron spin coherence exceeding seconds in high-purity silicon. *Nature Materials* **11**, 143-147, doi:Doi 10.1038/Nmat3182 (2012).
9   Simmons, M. Y. *et al.* Atomic-scale silicon device fabrication. *Int J Nanotechnol* **5**, 352-369 (2008).

10  Schmalbein, D., Maresch, G. G., Kamlowski, A. & Hofer, P. The Bruker high-frequency-EPR system. *Appl Magn Reson* **16**, 185-205 (1999).

11  Shtirberg, L. *et al.* High-sensitivity Q-band electron spin resonance imaging system with submicron resolution. *Rev. Sci. Instrum.* **82**, 043708, doi:10.1063/1.3581226 (2011).

12  Narkowicz, R., Suter, D. & Niemeyer, I. Scaling of sensitivity and efficiency in planar microresonators for electron spin resonance. *Rev. Sci. Instrum.* **79**, 084702 (2008).

13  Twig, Y., Suhovoy, E. & Blank, A. Sensitive surface loop-gap microresonators for electron spin resonance. *Rev. Sci. Instrum.* **81**, doi:Artn 104703:Doi 10.1063/1.3488365 (2010).

14  Twig, Y., Dikarov, E. & Blank, A. Ultra Miniature Resonators for Electron Spin Resonance: Sensitivity Analysis, Design and Construction Methods, and Potential Applications. *Molecular Physics - In press*, *avaiable on-line* (2013).

15  Twig, Y., Dikarov, E. & Blank, A. Cryogenic electron spin resonance microimaging probe. *J. Magn. Reson.* **218**, 22-29 (2012).

16  Twig, Y., Dikarov, E., Hutchison, W. D. & Blank, A. Note: High sensitivity pulsed electron spin resonance spectroscopy with induction detection. *Rev. Sci. Instrum.* **82**, 076105, doi:Artn 076105Doi 10.1063/1.3611003 (2011).

17  Rugar, D., Budakian, R., Mamin, H. J. & Chui, B. W. Single spin detection by magnetic resonance force microscopy. *Nature* **430**, 329-332 (2004).

18  Grinolds, M. S. *et al.* Nanoscale magnetic imaging of a single electron spin under ambient conditions. *Nat Phys* **advance online publication**, doi:http://www.nature.com/nphys/journal/vaop/ncurrent/abs/nphys2543.html#supplementary-information (2013).

19  Morton, J. J. L. *et al.* Solid-state quantum memory using the $^{31}$P nuclear spin. *Nature* **455**, 1085-1088, doi:Doi 10.1038/Nature07295 (2008).

**Figure captions**

Figure 1: The suggested QC scheme to be used in conjunction with ultra-high sensitivity/high-resolution induction detection[5]. A two-dimensional array of phosphorus atoms is produced inside a pure $^{28}$Si single crystal. The crystal is placed upside down on the center of our ultra-sensitive surface resonator (shown in the lower part of the figure)[14-16], and operated at cryogenic temperatures. Each phosphorus nucleus in the crystal serves as a logical quantum bit (qubit), while its adjacent electron is the working qubit. The array has two lattice constants: a short one (a) that enables electron spins to interact through dipolar couplings along this linear vector (similar to the manner described in ref [2]), and a long one (b) that separates many identical copies of the same individual vector computers. Individual spins can be addressed by applying a large magnetic field gradient with DC current into microwires (separating the spins in the frequency domain), and the state of all spins can be read out in parallel by a one-dimensional image along the crystal's x-axis. All parallel identical computer vectors should give the same vector of spin states, thereby increasing the measured signal and also greatly minimizing the need for quantum error correction due to ransom spin flips, since the measured result averages over ~100 spins per qubit. Information can be swapped between working electron spins and logical nuclear spins through combined radiofrequency (RF) and microwave (MW) pulse sequences, as described in reference[19].

Figure 2: ESR signal (blue lines) compared to noise level (red lines, obtained at 100-G off-resonance with 1-sec averaging time) for the $^{28}$Si:P sample placed on the 20-µm (left) and 5-µm (right) resonators. The two inserts show the noise level in millivolts (blown up by a factor of 1000).

Figure 3: Calculated and measured microwave magnetic field distribution ($B_1^2$) close to the resonator's surface. (a) Calculated $B_1^2$ on the 20-µm resonator, summed over the first 5 µm above the surface. (b) Two-dimensional ESR image taken with a flat $^{28}$Si:P sample placed on the resonator. (c) The same as in (a) but for the 5-µm resonator, summed over the first 3 µm above the surface (the area at the center of the resonator is blown up for better clarity). (d) The same as in (b) but for the 5-µm resonator.

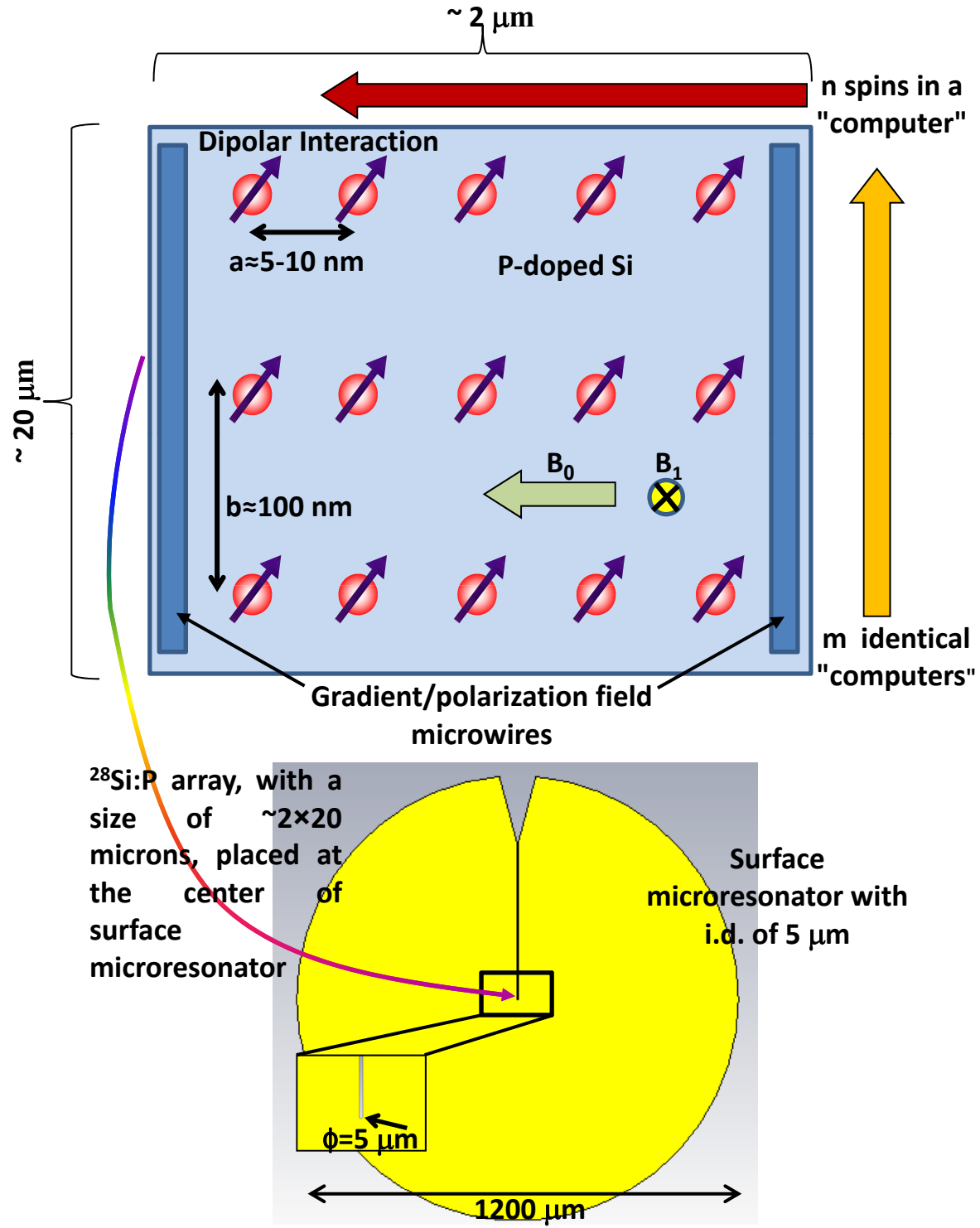

Figure 1

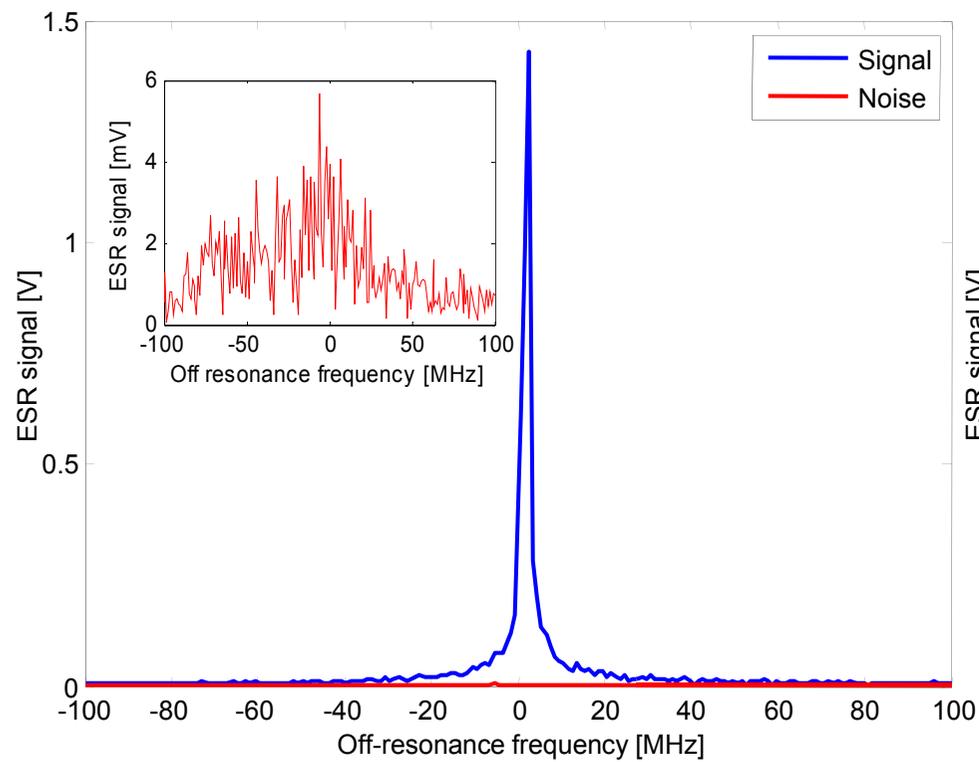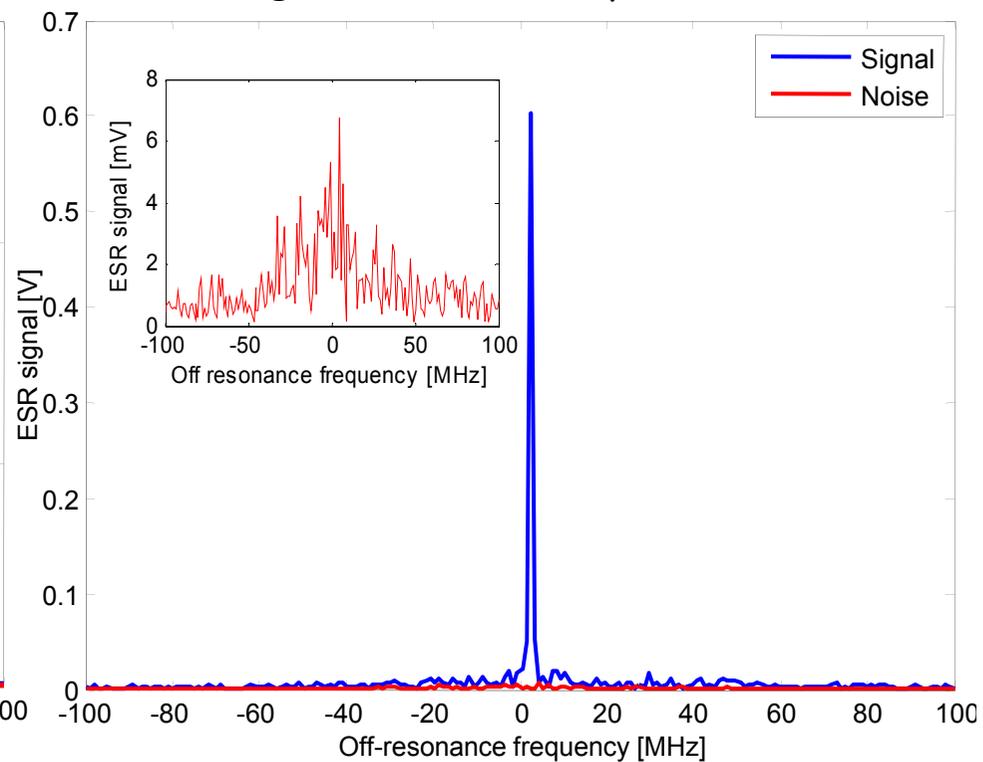

Figure 2

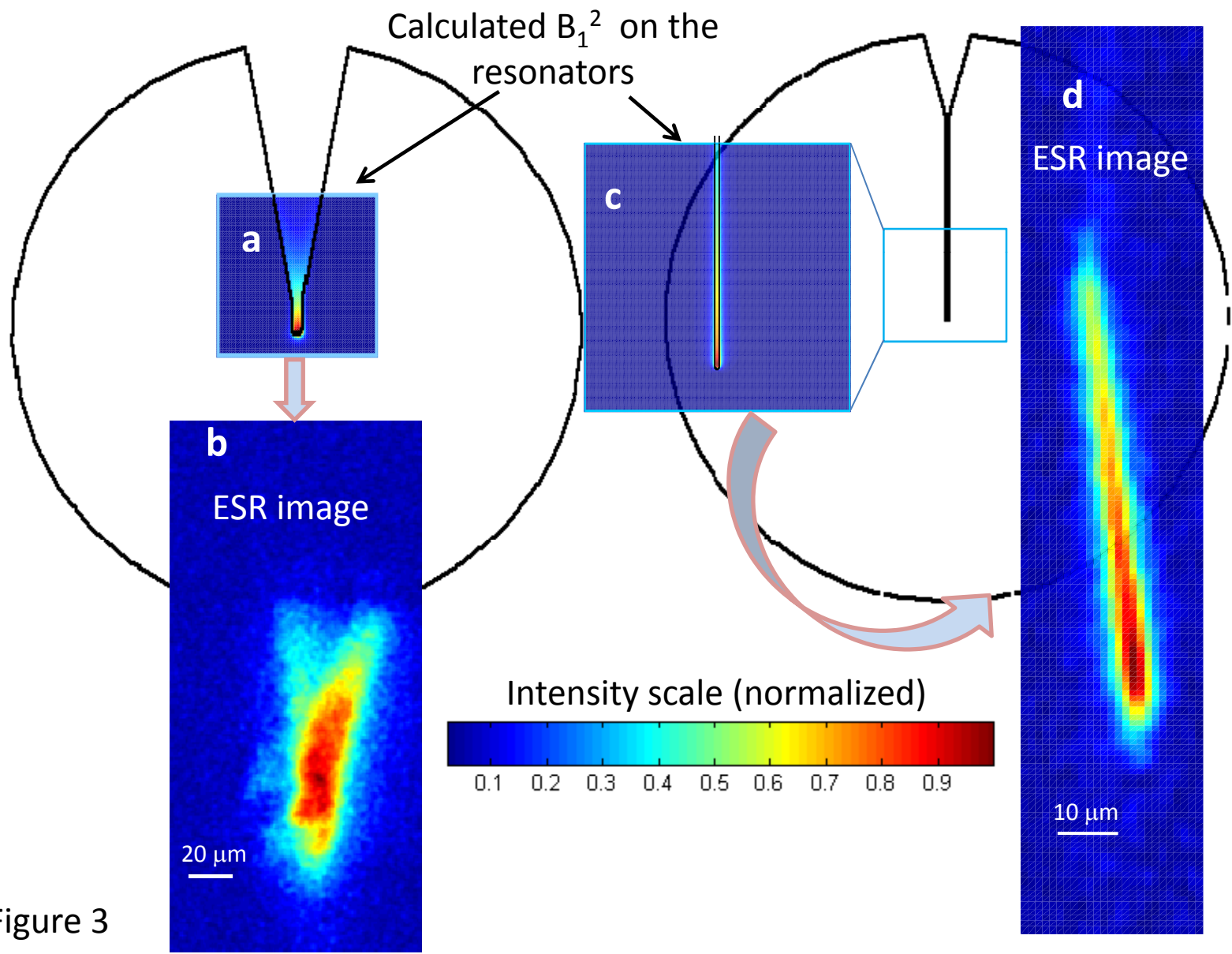

Figure 3

# Induction Detection Electron Spin Resonance with Sensitivity of 1000 Spins: En Route to Scalable Quantum Computations

Supplementary information

Figure S1: The new cryogenic probe that was employed in the experiments. The resonator is operated in Reflection mode. Both the circulator and the first low noise amplifier are cooled to cryogenic temperatures. The probe has several functionalities: (a) It facilitates the use of optical excitation by optical fiber, if needed. (b) It enables the generation of static and pulsed magnetic field gradients and polarization fields in all 3 axes. (c) It supports the use of current sensors for electrically-detected ESR experiments. (d) It has 2 (and in other designs, 3) independent piezo stages to control the coupling between the microwave line and the resonator. (e) Provision for independent temperature sensor reading.

Figure S2: Details of the microwave coupling configuration and the position of the sample with respect to the resonator. The image shows the 5-μmresonator, which uses a 20-μm resonator on the bottom part as an auxiliary resonator to facilitate efficient microwave energy coupling (see ref [14]).

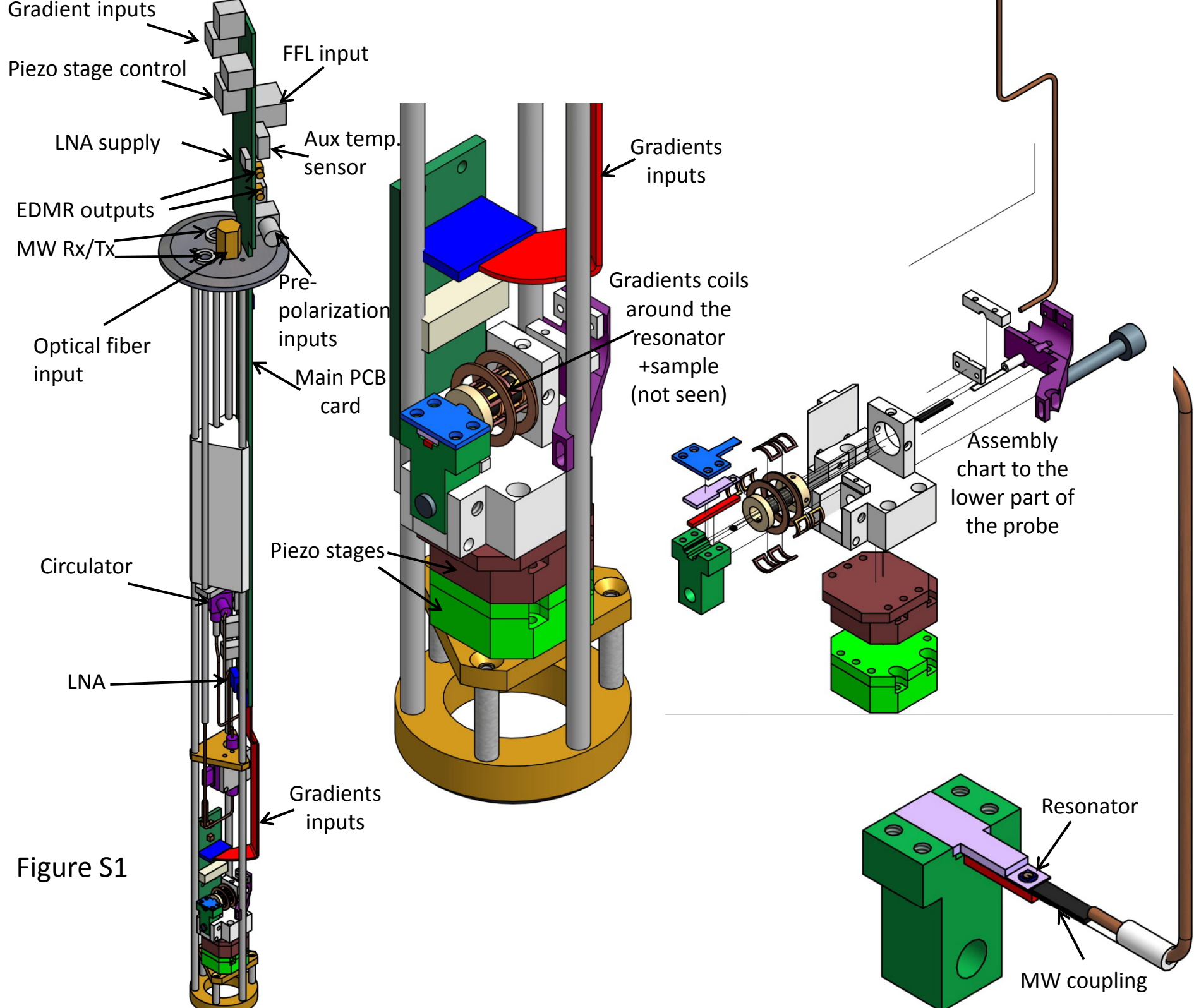

Figure S1

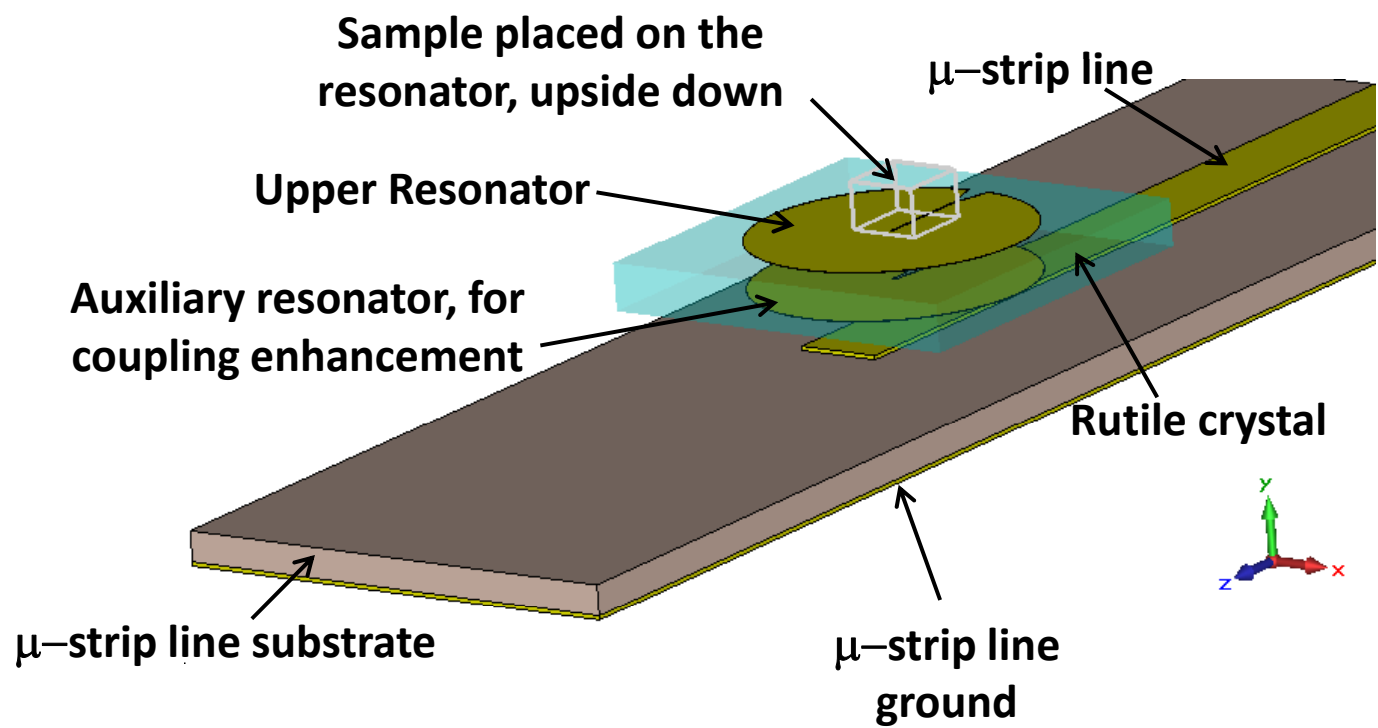

Figure S2